\documentclass[aps,pre,reprint,superscriptaddress]{revtex4-1}

\usepackage{amssymb,amsfonts,amsmath}
\usepackage{graphicx}
\usepackage{enumerate}
\usepackage[english]{babel}
\usepackage{siunitx}

\newcommand{\bareD}{D_0}
\newcommand{\Deff}{D_{\textit{\tiny{eff}}}}
\newcommand{\Deffs}[1]{D_{\textit{\tiny{eff}}\text{\tiny{,#1}}}}
\newcommand{\tts}[2]{{#1}_{\text{\tiny{#2}}}}  
\newcommand{\kbt}{{k}_{\text{\tiny{B}}}T}
\newcommand{\tMFPT}{{\tau}_{\text{\tiny{MFPT}}}}
\newcommand{\eg}{{\it e.g.\ }}
\newcommand{\ie}{{\it i.e.\ }}

\newcommand{\avg}[1]{\left\langle #1 \right\rangle}

\newcommand{\prth}[1]{\left( #1 \right)}
\newcommand{\brcs}[1]{\left\{ #1 \right\}}
\newcommand{\sqbr}[1]{\left[ #1 \right]}
\newcommand{\equn}{Eq.}
\newcommand{\fign}{Fig.}
\newcommand{\secn}{Sec.}
\renewcommand{\vec}[1]{\mathbf{#1}}
\newcommand{\figwidth}{0.48}

\makeatletter
  \renewcommand{\fnum@figure}{\textbf{Figure~\thefigure}}
  \makeatother

\begin{document}

\title{Diffusion in a rough potential revisited}

\author{Saikat Banerjee}
\affiliation{Solid State and Structural Chemistry Unit, Indian Institute of Science, Bangalore - 560012, India}
\author{Rajib Biswas}
\affiliation{Solid State and Structural Chemistry Unit, Indian Institute of Science, Bangalore - 560012, India}
\author{Kazuhiko Seki}
\affiliation{National Institute of Advanced Industrial Science and Technology (AIST), AIST Tsukuba Central 5 Higashi 1-1-1, Tsukuba, Ibaraki 305-8565, Japan}
\author{Biman Bagchi}
\email{bbagchi@sscu.iisc.ernet.in}
\affiliation{Solid State and Structural Chemistry Unit, Indian Institute of Science, Bangalore - 560012, India}
\vspace{1in}

\begin{abstract}
   Rugged energy landscapes find wide applications 
   in diverse fields ranging from astrophysics to protein folding. 
   We study the dependence of diffusion coefficient $(D)$ 
   of a Brownian particle on the distribution width 
   $(\varepsilon)$ of randomness in a Gaussian random landscape 
   by simulations and theoretical analysis. 
   We first show that the elegant expression of Zwanzig [PNAS, {\bf 85}, 2029 (1988)] for
   $D(\varepsilon)$ can be reproduced exactly by using the 
   Rosenfeld diffusion-entropy scaling relation.
   Our simulations show that Zwanzig's expression overestimates $D$ 
   in an uncorrelated Gaussian random lattice 
   -- differing by almost an order of magnitude at moderately high ruggedness.
   The disparity originates from the presence of 
   ``three-site traps'' (TST) on the landscape 
   -- which are formed by the presence of deep minima 
   flanked by high barriers on either side.
   Using mean first passage time formalism, 
   we derive a general expression for the effective 
   diffusion coefficient in the presence of TST,
   that quantitatively reproduces the simulation results 
   and which reduces to Zwanzig's form 
   only in the limit of infinite spatial correlation. 
   We construct a continuous Gaussian field with inherent 
   correlation to establish the effect of spatial
   correlation on random walk.
   The presence of TSTs at large ruggedness $(\varepsilon \gg \kbt)$
   give rise to an apparent breakdown of ergodicity of the type
   often encountered in glassy liquids.
\end{abstract}

\maketitle

\section{Introduction}
The diffusion of a Brownian particle on a random energy landscape 
serves as an effective model in understanding different complex phenomena 
and can be considered as a historically important~\cite{bagchi_molecular_relaxation,
lifson_jcp, weiss_math_phys,weiss_acp,zwanzig,frauenfelder_sci} problem.
Examples include diffusion in glassy matrices 
and supercooled liquids~\cite{stillinger_sci, angell_sci, ediger_jpc},
dynamics of molecular motors moving along heterogeneous substrates~\cite{nelson_biophys},
diffusion of a protein along a DNA in search for a specific binding site~\cite{blainey_nat},
dynamics of fluorescently labeled molecules inside the cell~\cite{metzler_prl, edward_golding}.
A highly topical application of this model is in protein folding where the transformation of
the unfolded state is viewed as diffusion in polymer conformation space that contains
multiple maxima and minima~\cite{dill_biochem, bryngelson_wolynes_1, bryngelson_wolynes_2, dill_2}.
Yet another example is provided by enzyme kinetics where a broad distribution 
of relaxation times observed in single molecule spectroscopy has been 
attributed to random energy landscape experienced by the enzyme 
near the global minimum that determines its 
equilibrium configuration~\cite{xie_ncb,xie_acr}. 
For many years different variant of random energy barrier models 
have been used to study electron transport in disordered solids~\cite{scher_prb}.

Despite the broad applicability and historical importance of the problem, 
there are surprisingly few numerical and simulation studies of this problem. 
As a result, we have little knowledge about the effect of 
ruggedness on diffusion at a quantitative level. 
Every study seems to use the expression of Zwanzig (discussed below) 
but the validity of the same has never been tested, 
although Zwanzig himself termed his derivation as ``conjectural''. 
There have been studies on random traps and random barriers, 
but true definition of ruggedness requires simultaneous presence of both. 
Diffusion in rugged landscape is thus quantitatively
different from either random traps or random barriers.

Another important issue not touched upon adequately is 
the role of spatial correlations in the diffusion process.
Ruggedness is expected to be correlated in many cases, 
such as protein diffusion along a DNA.
Such correlations can alter the motion of a particle.
Wolynes and co-workers have shown that the dynamics
changes considerably in presence of correlations in 
protein folding funnels~\cite{shoemaker_wolynes_wang_pnas,plotkin_wang_wolynes_jcp} 
and glass transitions~\cite{wolynes_jphysfrance_1997}.

The dynamics of a free Brownian particle at time scales 
where inertia can be neglected is well understood
since Einstein's seminal paper, 
and has been generalized in many different directions~\cite{klafter_phys_rep}.
However, in the presence of a random potential 
with multiple maxima and minima, 
diffusion can become significantly different 
as the simultaneous presence of barriers and troughs can
significantly and non-trivially retard the mean square displacement. 
Several models have been developed to understand the complex dynamics, 
\eg the random trap model~\cite{kehr_prb},
the random barrier model~\cite{alexander_prl, sollich_stat_mech},
continuous time random walk~\cite{angell_sci}, etc.
Theoretical analyses are mostly restricted to asymptotic long-time limits, 
when the particle motion should become diffusive. 
In an important treatment of the problem, Zwanzig~\cite{zwanzig}
considered a general rough potential ${U(x)}$ with a smooth background ${U_{0}(x)}$
on which a perturbation ${U_{1}(x)}$ is superimposed, so that $U(x) = U_{0}(x) + U_{1}(x)$
He showed that the effective diffusion coefficient $(\Deff)$ on the rough potential can be expressed as
\begin{equation} 
\label{eq:D_zwanzig_main}
    \Deff =\frac{\bareD}{\avg{e^{\beta U_{1}}} \avg{e^{-\beta U_{1} }} }  
\end{equation}
where $\bareD$ is the bare diffusion coefficient on the smooth potential, 
$\beta = 1/\kbt$ and $\avg{\ldots} $ denotes the spatial, \emph{local} average
used to smooth the perturbation.
For a random potential, where the amplitude of roughness has a Gaussian distribution,
\begin{equation} 
\label{eq:gauss_prob}
    P\prth{U_{1}} =\frac{1}{\varepsilon \sqrt{2\pi }} \sqbr{\exp \prth{-\frac{U_{1}^{2} }{2\varepsilon ^{2} }}} 
\end{equation}
in which $\varepsilon$ is the root-mean-squared roughness, 
$\varepsilon ^{2} = \avg{U_{1}^{2}}$,
Zwanzig showed that the effective diffusion coefficient $\Deffs{Z}$
can be given by the following simple and elegant expression,
\begin{equation} 
\label{eq:D_zwanzig_gauss}
    \Deffs{Z} = \bareD \exp \prth{-\beta ^{2}\varepsilon ^{2}} 
\end{equation}
Note that we use the subscript on $\Deffs{Z}$ to refer Zwanzig's work,
and will use different subscripts as we discuss 
further for the sake of comparison and analyses.
Despite the novelty of the work and simplicity of the expression, 
the derivation of the above invokes the questionable 
local averaging of the random energy surface 
(in the simplification of the double integral that arises while evaluating the MFPT).
\emph{Zwanzig himself was aware of the possible limitation of his approximate approach, 
and termed his final result as ``conjectural''}.

\begin{figure}
\begin{center}
  \includegraphics[width=\figwidth\textwidth]{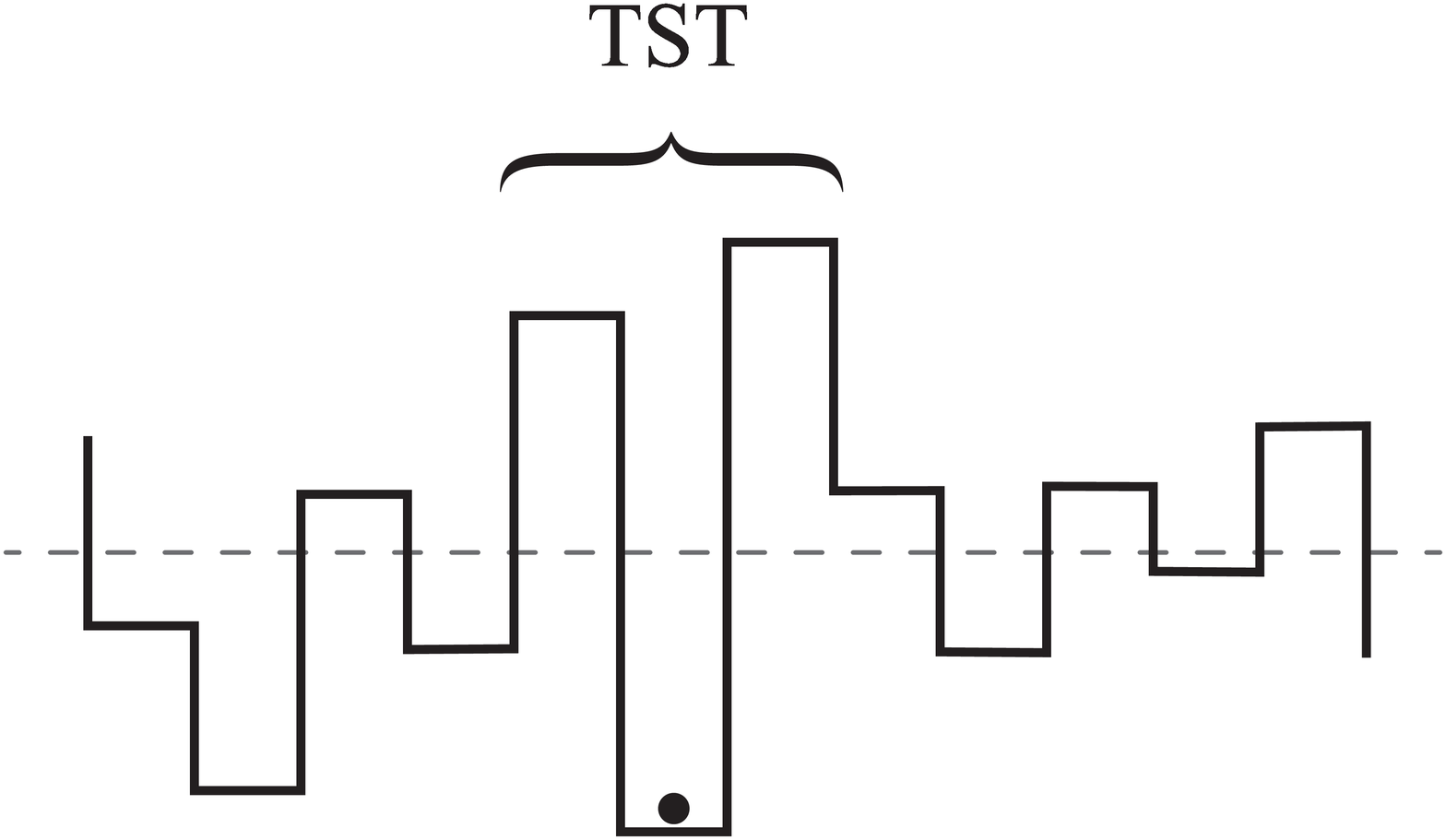}
\end{center}
\caption
      { \label{fig:diff_tst}
        Schematic representation of ``Three-Site Trap'' (TST) on a lattice (not to scale).
        Unlike isolated extremely deep minimum or maximum, TST is formed
        when a deep minimum is flanked by high maxima on both sides.
        The probability of encountering a TST would be the
        joint probability of occurrence of such three sites simultaneously.
        Hence TSTs are more common than extremely sharp minima / maxima
        in a Gaussian random surface.
      }
\end{figure}

There are multiple unanswered issues in this problem.
First and foremost, the existence of diffusion itself 
could be doubtful at large ruggedness.
Imagine that the particle encounters a situation 
where it is stuck in a deep minimum (negative energy)
with maxima (barriers, positive energy) on its two sides. 
We refer to this as ``three-site trap (TST)'' (see \fign~\ref{fig:diff_tst}).
Such TSTs become increasingly probable as ruggedness 
(that is, $\varepsilon $ ) increases,
and can give rise to long trapping 
and hence sub-diffusive growth of MFPT.
In fact, such TSTs are ignored in the coarse-graining mentioned earlier.
We find that exact evaluation of the MFPT 
deviates from and improves upon the coarse-grained expression.
Second, the MFPT approach to estimate the diffusion constant 
(comparing $\tMFPT$ with that of an effective flat potential, 
thereby implicitly invoking the relation $\Deff = L^{2} / 2\tMFPT$
where $\tMFPT$ is the MFPT between an initial and 
final position separated by a distance ${L}$) 
might not work. 
Third, and a related issue, is the question of stationarity and ergodicity.
Diffusion can be defined for a random process which is both stationary and ergodic.
Even if we consider a random landscape which is stationary,
the long trapping in the deep minima results in a ``broken ergodicity'' 
on such random potential energy surface, 
which has strong resemblance with the glass transition scenario. 
Modeling motion on such landscapes using standard methods of Monte Carlo is unreliable,
and requires special asymptotic techniques~\cite{wang_spiliopoulos_2011, wang_spiliopoulos_2012}.
This paradigm for trapping on long timescales by metastable states 
in complex systems may be visualized as a terrain with lakes in the valleys whose
water level depends on the observational timescale~\cite{stein_prob_stat}. 
This is where the relationship between diffusion and entropy can have a role to play. 
Last but not the least, there could be a spatial
correlation among the values of energy of the neighboring sites. 
Such a correlation adds a new dimension to the problem. 
Fortunately, we have been able to address all the four issues in this work.

\section{Rosenfeld entropy scaling on rugged energy landscape}
The successful entropy-diffusion scaling relationship
was first proposed by Rosenfeld in 1977 on the 
basis of extensive simulation results 
for the transport coefficients of a wide variety of 
one-component systems including those containing hard spheres, 
soft spheres, or plasma~\cite{rosenfeld_cpl, rosenfeld_pra}.
Using macroscopic reduction parameters for the length as $\rho ^{-{1/3}} $ 
and the thermal velocity as $\left({\kbt / m}\right)^{1/2}$, 
Rosenfeld demonstrated that, in dimensionless units,
the self-diffusivity $\bareD$ of a bulk fluid is well 
correlated with the excess entropy in terms of
an exponential relation
\begin{equation} 
\label{eq:rosenfeld_scaling}
    \Deffs{R} = \bareD \frac{\rho^{1/3}}{\prth{{\kbt/m}}^{1/2}} \approx a\exp\prth{b\tts{S}{ex}}
\end{equation}
where $\tts{S}{ex} = (S-\tts{S}{id}) / N\tts{k}{B}$ is the reduced
excess (dimensionless) entropy per molecule, 
$a$ and $b$ are the constants which depend on the system,
but $b$ shows weak variation. 
Although Rosenfeld scaling relation is routinely used in 
varied contexts for understanding the relationship 
between thermodynamics, transport properties and 
potential energy landscape ~\cite{charusita_agarwal_pre_2008, charusita_jcp_2008},
the validity of the relation was
established by essentially empirical means.
Another well-known relationship between entropy and diffusion 
in glassy liquids was given by Adam and Gibbs~\cite{adam_gibbs_1}, 
and is of the following form,
\begin{equation} 
\label{eq:adam_gibbs_scaling}
    \Deffs{AG} =a\exp \prth{-\frac{b}{T\tts{S}{conf}}}
\end{equation}
where $\tts{S}{conf}$ is the configurational entropy. 
In the intermediate temperature regime,
both Rosenfeld and Adam-Gibbs seem to provide reliable descriptions, 
although at still lower temperature, in viscous liquid, 
Rosenfeld scaling becomes unreliable. 
Surprisingly, relationship between these
two entropy-based relations has not been sufficiently explored.

A random energy landscape with Gaussian distribution
allows an exact derivation of partition function,
which leads us to the excess entropy.
The connection between entropy and random energy landscapes 
was earlier discussed by Wolynes~\cite{wolynes_entropy} 
Such correlation helps us to connect the Rosenfeld scaling relation
with Zwanzig's diffusion coefficient.
Therefore, it provides an indirect way to theoretically validate
the Rosenfeld scaling relation.
Starting with the partition function $(Q)$ for the
random energy surface, we obtain free energy $(A)$ and entropy $(S)$ as,
\begin{equation} 
\label{eq:rosenfeld_deriv_Q}
    Q = \sum_{i} \exp \prth{-\frac{U_{1i}}{\kbt}} = N\exp \prth{\frac{\varepsilon^{2}}{2(\kbt)^2}}
\end{equation}
\begin{equation} 
\label{eq:rosenfeld_deriv_A}
    A = -\kbt \ln Q = -\kbt \ln N - \frac{\varepsilon^{2}}{2 \kbt}  
\end{equation}
\begin{equation} 
\label{eq:rosenfeld_deriv_S}
    S = -\prth{\frac{dA}{dT}} = \tts{k}{B} \ln N - \frac{\varepsilon^{2}}{2\kbt^{2}}  
\end{equation}
where $\tts{k}{B} \ln N$ is the ideal gas contribution. 
Hence the excess (dimensionless)
entropy $\tts{S}{ex}$ (defined by Rosenfeld), 
for a single particle becomes,
\begin{equation} 
\label{eq:rosenfeld_deriv_Sex}
    \tts{S}{ex} = \frac{S - \tts{S}{id}}{\tts{k}{B}} = -\frac{\varepsilon^{2}}{2(\kbt)^2}  
\end{equation}
from which we obtain the effective diffusion coefficient,
\begin{equation} 
\label{eq:rosenfeld_deriv_D}
    \Deffs{R} = a \exp \prth{-\frac{b}{2} \beta^{2} \varepsilon^{2}}
\end{equation}
which, in essence, is equivalent to Zwanzig's 
expression [\equn~\ref{eq:D_zwanzig_gauss}].
By comparing \equn~\ref{eq:D_zwanzig_gauss} and~\ref{eq:rosenfeld_deriv_D}, 
we obtain the Rosenfeld scaling parameters as $a=D_{0}$, $b=2$. 
The value of $b$ is close to the values reported for this constant.

\section{Model I. Gaussian discrete lattice}

\subsection{Description of the model}

\begin{figure}
\begin{center}
  \includegraphics[width=\figwidth\textwidth]{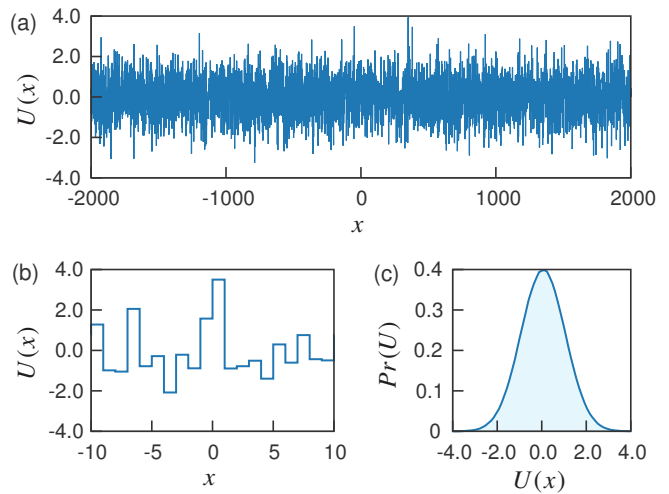}
\end{center}
\caption
      { \label{fig:diff_discrete_lattice_model}
        (a) Discrete random Gaussian potential at $\varepsilon =1.0$. 
        (b) Zoomed-in portion of the potential showing discrete 
            lattice sites forming barriers and traps.
        (c) Distribution of the potential energy 
            at each lattice site showing Gaussian behavior.
      }
\end{figure}

We introduce a discrete random lattice, 
where the energy of each site is sampled from a 
Gaussian distribution of mean zero and variance $\varepsilon$. 
A similar but different model of random traps and barriers 
was earlier introduced by Limoge and Bocquet~\cite{bocquet_prl}, 
and later studied by Kehr and co-workers~\cite{kehr_cond_mat}. 
To contrast, the earlier model had strictly alternating barriers and traps,
with a restriction of positive values on barrier energies 
and negative values on trap energies.
Transitions were allowed only from one trap 
to the next, crossing the barrier. 
Our model does not restrict the energy values at individual lattice sites. 
Hence there can be three different scenarios,
\begin{enumerate}[(i)]
  \item a lower energy site neighbored by two higher
        energy sites (trap with barrier on both sides)
  \item a higher energy site neighbored by lower energy sites
        (barrier with trap on both sides) 
  \item a site neighbored by higher energy on one side and lower
        energy on other side (barrier on one side, trap on other side)
\end{enumerate} 
The random walker is allowed to visit any of the neighboring sites, 
irrespective of the site behaving as a barrier or trap.  
The first case is of special interest, 
and we have termed it as ``three-site trap (TST)'' (see \fign~\ref{fig:diff_tst}). 
In \fign~\ref{fig:diff_discrete_lattice_model}(a), we show
the discrete random potential at $\varepsilon =1.0$.
The random potential consists of discrete lattice sites 
[see \fign~\ref{fig:diff_discrete_lattice_model}(b)]
with the energy at each lattice site sampled from a Gaussian distribution
[see \fign~\ref{fig:diff_discrete_lattice_model}(c)].

For particle diffusion on this potential, 
it seems reasonable to restrict the transitions to nearest neighbors.
All transitions to neighbor sites have identical rates, 
if the final site has a lower energy than the initial site.
Transitions that lead to energetically higher sites require thermal activation,
\begin{equation} 
\label{eq:discrete_transition_prob} 
  \Gamma_{i,j} = 
      \begin{cases}
        \Gamma_{0}                                          & U_{j} < U_{i}   \\
        \Gamma_{0} \exp \sqbr{-\beta\prth{U_{j} -U_{i}}}    & U_{j} \ge U_{i}
      \end{cases}  
\end{equation}
where $\Gamma_{i,j} $ is the transition rate 
from site ${i}$ to ${j}$ and $\beta = 1 / \kbt$.
Transitions of this type were earlier introduced 
by Miller and Abrahams~\cite{miller_phys_rev}.
We perform a continuous time random walk (CTRW) 
on this potential, assuming $\Gamma_{0} =1$ .
The random walker at any site can move 
either to the left or to the right with rates, 
$\Gamma_{l} $ and $\Gamma_{r} $ respectively. 
We call a random number $(r)$ to decide 
the move to the left or right with 
probabilities ${\Gamma_{l} / \Gamma_{tot} }$ and ${\Gamma_{r} / \Gamma_{tot} }$ 
where, $\Gamma_{tot} =\Gamma_{l} +\Gamma_{r} $.
The time required for the move in the CTRW is given as,
\begin{equation} 
\label{eq:ctrw_time}
  \Delta t=-\frac{\ln r}{\Gamma_{tot} }
\end{equation}

\subsection{Simulation results}

\begin{figure}
\begin{center}
  \includegraphics[width=\figwidth\textwidth]{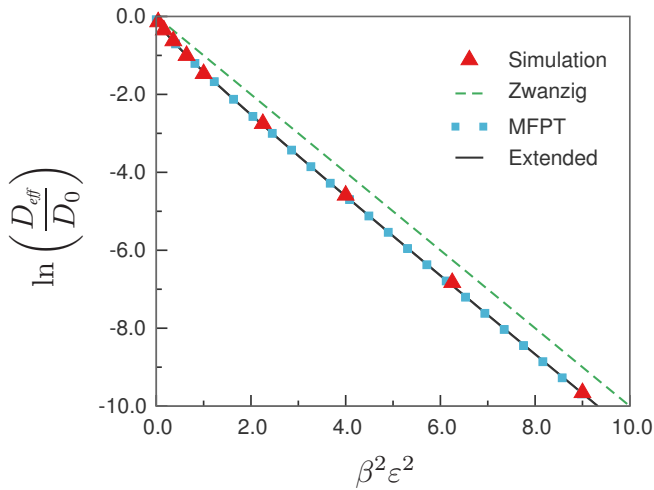}
\end{center}
\caption[]
      { \label{fig:diff_discrete_lattice_result}
        Semilog plot of the scaled self-diffusion coefficient $\Deff$
        against the squared ruggedness parameter $(\varepsilon)$ 
        for the discrete Gaussian random lattice. 
        The dashed green line represents the theoretically
        predicted values of Zwanzig [\equn~\ref{eq:D_zwanzig_gauss}]. 
        The solid red triangles represent the results of CTRW simulation. 
        The result obtained from numerical evaluation of MFPT 
        [Eqs.~\ref{eq:tMFPT_lattice_db} and \ref{eq:D_mfpt_lattice_we}] 
        on the potential surfaces is indicated by blue squares. 
        The solid black line shows the diffusion coefficient obtained from the 
        corrected equation [\equn~\ref{eq:D_lattice_final}].
      }
\end{figure}

We perform CTRW of a Brownian particle on the discrete lattice
for \num{1e7} steps. 
The mean-square displacement $\avg{\Delta x^{2}}$ of the random walker
gives us the effective diffusion coefficient following Einstein's relation,
\begin{equation} 
\label{eq:D_einstein}
  \avg{\Delta x^{2}} = 2\Deff t 
\end{equation}
where $\Deffs{S}$ denotes the $\Deff$ obtained from simulation.
The observed $\Deffs{S}$ with varying randomness $(\varepsilon)$ 
is compared with the theoretically predicted values of 
Zwanzig [\equn~\ref{eq:D_zwanzig_gauss}] 
in \fign~\ref{fig:diff_discrete_lattice_result}.
Even at small $\varepsilon$, within the well-defined diffusive limit, 
Zwanzig's prediction is found to systematically overestimate
(by a small but non-trivial amount) the simulated diffusion coefficient. 
\emph{Deviation from Zwanzig's expression becomes large by 
$\varepsilon =3.0$ where the former overestimates 
the simulated value by almost an order of magnitude}. 
In contrast, the estimate of $\Deffs{MFPT}$ from the exact numerical
evaluation of MFPT (see Eqs.~\ref{eq:tMFPT_lattice_db} 
and~\ref{eq:D_mfpt_lattice_we} below) provides 
quantitative agreement with the simulation results. 
The corrected expression of the diffusion coefficient 
(see \equn~\ref{eq:D_lattice_final} below) that we have derived,
also provides a quantitative agreement.

\subsection{Numerical analysis from MFPT}

To understand the deviation from Zwanzig's prediction, 
we looked into the validity of coarse-graining of the 
potential energy surface in his derivation. 
Without averaging over the random surface, 
it is possible to derive the self-diffusion
coefficient using MFPT. 
On a segment of linear chain with $N+1$ sites, with a reflecting
boundary condition at site 0 and an absorbing boundary condition at site $N$, 
an exact expression is known~\cite{kehr_pra}
for MFPT $(\tMFPT)$, for fixed disorder in which all transition rate for the
segment appear explicitly,
\begin{equation} 
\label{eq:tMFPT_lattice_main} 
    \tMFPT =\sum_{i=0}^{N-1}\frac{1}{\Gamma_{i,i+1}} 
          + \sum_{i=1}^{N-1}\frac{1}{\Gamma_{i,i+1}} \sum_{j=0}^{i-1}\prod_{k=j}^{i-1}\frac{\Gamma_{k+1,k}}{\Gamma_{k,k+1}}    
\end{equation}
We can consider detailed balance to be valid between two neighboring sites,
\begin{equation}
\label{eq:detailed_balance}
    \rho_{i} \Gamma_{i,j} = \rho_{j} \Gamma_{j,i} \mbox{   with } 
    \rho_{i} = \frac{\exp \prth{-\beta U_{i}}}{\brcs{\exp \prth{-\beta U_{i}}}}
\end{equation}
where $\{\ldots\}$ denote the disordered average, 
and the neighboring sites are given by ${i}$ and ${j}$.
Here, $\rho_{i}$ is an occupation factor, 
which is proportional to the occupation probability of site ${i}$.
Introducing the detailed balance condition in 
\equn~\ref{eq:tMFPT_lattice_main} we obtain,
\begin{eqnarray} 
\label{eq:tMFPT_lattice_db}
  \tMFPT &=& \sum_{i=0}^{N-1} \frac{1}{\Gamma_{i,i+1} }  
            +\sum_{i=1}^{N-1}\sum_{j=1}^{i} \frac{\rho_{j-1}}{\rho_{i}} \frac{1}{\Gamma_{i,i+1} } \nonumber\\  
         &=& \sum_{i=0}^{N-1} \frac{1}{\Gamma_{i,i+1} }
            +\sum_{i=1}^{N-1}\sum_{j=1}^{i} \frac{\exp \sqbr{-\beta \prth{U_{j-1} -U_{i}}}}{\Gamma_{i,i+1}} \nonumber\\
         & &
\end{eqnarray}
\equn~\ref{eq:tMFPT_lattice_db} gives an exact expression 
for the MFPT on a discrete lattice under equilibrium conditions.
We calculate the MFPT on our quenched discrete potential 
by explicitly evaluating the summations numerically, 
and obtain the diffusion coefficient using the asymptotic relation,
\begin{equation} 
\label{eq:D_mfpt_lattice_we}
    \Deffs{MFPT} = \mathop{\lim}\limits_{N \to \infty } \frac{N^2}{2\tMFPT}  
\end{equation}
It is to be noted here that \equn~\ref{eq:D_mfpt_lattice_we} 
assumes that the rough energy surface 
can be replaced by an effective flat energy surface. 
As shown in \fign~\ref{fig:diff_discrete_lattice_result}, 
the numerical evaluation of 
$\Deffs{MFPT}$ provides quantitative agreement 
with the results of CTRW on the discrete potential. 
We note that the MFPT explicitly
takes into account the effect of TSTs, 
wherein a very deep trap is neighbored by two maxima.
This would be neglected if one does a coarse-grained average, 
as in Zwanzig's treatment. 
Probability of occurrence of such deep traps increases 
with increasing randomness.

\subsection{Theoretical derivation}

With the success of numerical analysis using MFPT,
one would expect a correct analytical expression 
for $\Deff$ derived with MFPT formalism.
Here we show the derivation of an elegant analytical expression for $\Deff$.
We start with \equn~\ref{eq:tMFPT_lattice_db},
which on further simplification gives,
\begin{equation} 
    \tMFPT = \sum_{i=0}^{N-1}\sum_{j=0}^{i} \frac{\exp \sqbr{-\beta \prth{U_{j} -U_{i}}}}{\Gamma_{i,i+1}}
\end{equation}
With no loss of generality for the system under translational invariance, 
we can do an averaging over the potential,
\begin{equation} 
\label{eq:tMFPT_lattice_avg} 
    \tMFPT = \sum_{i=0}^{N-1}\sum_{j=0}^{i} \avg{\frac{\exp \sqbr{-\beta \prth{U_{j} -U_{i}}}}{\Gamma_{i,i+1}}}
\end{equation}
where and below the same symbol is used for the mean first passage time and its ensemble average.
By introducing the transition rate of the Miller-Abraham process, 
given by \equn~\ref{eq:discrete_transition_prob} 
in \equn~\ref{eq:tMFPT_lattice_avg}, we obtain,
\begin{widetext}
\begin{align} 
\label{eq:tMFPT_lattice_MA}
   \tMFPT &= \frac{1}{\Gamma_0} 
             \sum_{i=0}^{N-1} \sum_{j=0}^{i} \brcs{ \avg{ \exp \sqbr{ -\beta \prth{U_{j} -U_{i+1}}}}_{U_{i+1} \ge U_{i}}
                                                  + \avg{ \exp \sqbr{ -\beta \prth{U_{j} -U_{i}  }}}_{U_{i+1} <   U_{i}} }
             \nonumber\\
          &= \frac{1}{\Gamma_0} \brcs{ \sum_{i=0}^{N-1} \sum_{j=0}^{i} + \sum_{i=-1}^{N-2}\sum _{j=0}^{i-1}}
                                      \avg{ \exp \sqbr{-\beta \prth{U_{j} -U_{i+1}}}}_{U_{i+1} \ge U_{i}}
\end{align}
In the limit of $N\gg 1$ and in the absence of spatial correlation, 
we can simplify \equn~\ref{eq:tMFPT_lattice_MA} to obtain,
\begin{equation} 
\label{eq:tMFPT_lattice_MA_avg} 
    \tMFPT = \frac{2}{\Gamma_0} \sum_{i=0}^{N-1} \sum_{j=0}^{i} 
             \avg{ \exp \prth{-\beta U_{j  }}} 
             \avg{ \exp \prth{ \beta U_{i+1}}}_{U_{i+1} \ge U_{i}}
\end{equation}
The average can be taken by using the Gaussian probability distribution,
\begin{equation} 
\label{eq:avg_to_int}
    \avg{ \exp \prth{ \beta U_{i+1}}}_{U_{i+1} \ge U_{i}} = 
    \int\limits_{0}^{\ \infty} d\Delta U_{i} \int\limits_{-\infty }^{\ \infty } dU_{i} 
    \exp \sqbr{ \beta \prth{U_{i} + \Delta U_{i}}} 
    P\prth{U_{i}} P\prth{U_{i} +\Delta U_{i}}
\end{equation}
\end{widetext}
where $\Delta U_{i} = U_{i+1} - U_{i} $. 
Note that $\Delta U_{i}$ is independent of $U_{i}$ in the absence of spatial correlations. 
The effect of spatial correlations will be studied in \secn~\ref{sec:spatial_correlation}.
After straightforward calculation one obtains,
\begin{equation}
\label{eq:lattice_int_result}
    \avg{ \exp \prth{ \beta U_{i+1}}}_{U_{i+1} \ge U_{i}} =
        \frac{1}{2} \exp \prth{ \frac{\beta^{2}\varepsilon^{2}}{2}}
        \sqbr{ 1+{\rm erf} \prth{ \frac{\beta \varepsilon}{2} }} 
\end{equation}
where the right-hand side is independent of the index $i$ owing to the translational invariance.
Hence, using \equn~\ref{eq:lattice_int_result} in 
\equn~\ref{eq:tMFPT_lattice_MA_avg}, we obtain,
\begin{equation} 
\label{eq:tMFPT_lattice_final}
    \tMFPT = \frac{N^2}{2\Gamma_0} \exp \prth{\beta^{2} \varepsilon^{2}}
             \sqbr{ 1+{\rm erf} \prth{ \frac{\beta \varepsilon}{2} }}
\end{equation}
In a very different context, 
this type of equation was obtained earlier~\cite{cordes_prb}.
Using \equn~\ref{eq:tMFPT_lattice_final} and 
\ref{eq:D_mfpt_lattice_we}, 
we get the expression for diffusion coefficient as,
\begin{equation} 
\label{eq:D_lattice_final} 
    \Deff = \bareD \exp \prth{-\beta^{2} \varepsilon^{2}}
            \sqbr{ 1+{\rm erf} \prth{ \frac{\beta \varepsilon}{2} }}^{-1}
\end{equation}
The corrected diffusion coefficient 
improves upon the Zwanzig's expression and 
quantitatively agrees with
the simulation results (see \fign~\ref{fig:diff_discrete_lattice_result}).

\section{Model II. Gaussian random field}
\begin{figure}
\begin{center}
  \includegraphics[width=\figwidth\textwidth]{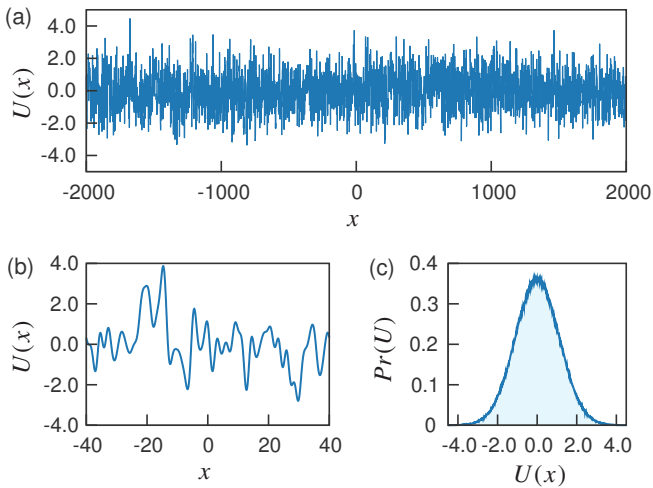}
\end{center}
\caption[]
      { \label{fig:diff_gaussian_field_model}
        (a) Continuous random Gaussian field [\equn~\ref{eq:gaussian_field}] at $\varepsilon =1.0$.
        (b) Zoomed-in portion of the potential showing the 
            continuity of the surface.
        (c) Distribution of the potential energy
            showing Gaussian behavior.
      }
\end{figure}

Our second model comprise of a continuous 
Gaussian random surface (or, field) [$\Phi $] 
generated by random Fourier modes. 
Using a standard method~\cite{kraichnan_fluid_mech},
we write the continuous random field as
\begin{equation} 
\label{eq:gaussian_field} 
    \Phi = \varepsilon \sqrt{\frac{2}{M}} \sum_{n=1}^{M} \cos \prth{ \vec{k}_{n} \cdot x + \theta_{n}}  
\end{equation}
where ${M}$ is the number of modes chosen, 
$\vec{k}_{n}$ is a random wave vector 
chosen independently from a Gaussian distribution 
of mean zero and variance $\sigma $ (we use $\sigma =1$),
$\theta _{n} $ is a random phase chosen from a 
uniform distribution between $0$ and $2\pi $. 
It can be shown that $\Phi $ has a Gaussian
distribution with mean zero and variance $\varepsilon $. 
In \fign~\ref{fig:diff_gaussian_field_model}, 
we show a realization of continuous random surface at
$\varepsilon =1.0$.
It has been generated using $200$ random modes $\left(M=200\right)$.
At smaller length scales, one can note the
continuity of the potential \fign~\ref{fig:diff_gaussian_field_model}(b).
The distribution of the potential energy of this lattice 
is shown in \fign~\ref{fig:diff_gaussian_field_model}(c).
Study of random walks on Gaussian random fields
has generated a lot of interest in recent years~\cite{olla_siri,remi_2009_01,remi_2009_02,wang_spiliopoulos_2012}.
The continuity of such a field helps us to perform continuous
Brownian Dynamics (BD), thereby providing an opportunity to 
probe the detailed dynamics of the system.
 
\begin{figure}
\begin{center}
  \includegraphics[width=\figwidth\textwidth]{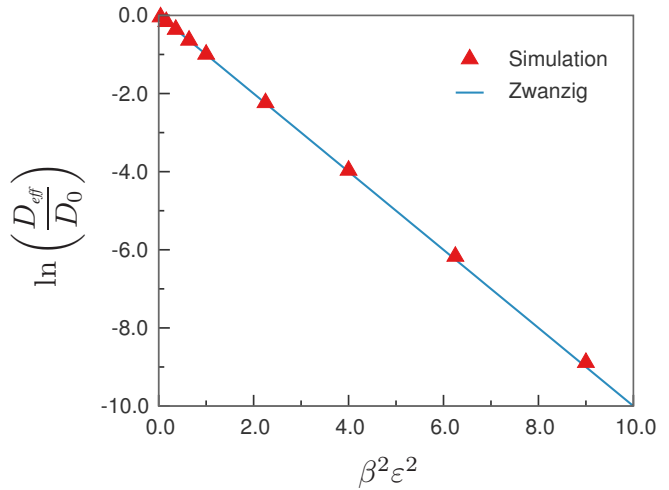}
\end{center}
\caption[]
      { \label{fig:diff_gaussian_field_result}
        Semilog plot of the scaled self-diffusion coefficient $\Deff$
        against the squared ruggedness parameter $(\varepsilon)$
        for the continuous Gaussian random field.
        The solid blue line shows the theoretically 
        predicted values of Zwanzig [\equn~\ref{eq:D_zwanzig_gauss}], 
        while the solid red triangles are the results of simulation.
      }
\end{figure}

We perform BD (using second-order Runge-Kutta method) 
on this Gaussian field with 1000 particles starting from random positions.
The effective diffusion coefficient obtained from the 
Brownian Dynamics simulation is compared with the
Zwanzig's expression in \fign~\ref{fig:diff_gaussian_field_result}.
Contrary to the discrete model, the simulation results 
corroborates with Zwanzig's expression. 
As we show below, the \emph{surprising} agreement 
is because of the nature of
the potential surface. 
The three-site traps (TSTs), 
which were increasingly dominant in the discrete model,
become negligible due to the inherent correlation in the continuous potential surface.
\begin{equation} 
\label{eq:field_correlation}
    \avg{\Phi(x)\Phi(x+\Delta x)} = \varepsilon^2 \exp\prth{-\frac{\sigma^2(\Delta x)^2}{2}}
\end{equation}
Due to the presence of this spatial correlation, 
the correction term for the continuous potential becomes negligible.

\section{Role of spatial correlation}
\label{sec:spatial_correlation}
A major motivation of the present work is to investigate 
the role of spatial correlations in the energy landscape on the
self-diffusion coefficient.
Examples of such correlations are abundant in nature. 
For example, correlations are known~\cite{stanley_nature_1992, 
stanley_pre_1995, buldyrev} 
to be present in the DNA sequence that a protein experiences
during search for its specific binding site. 
Similar correlations are also present in the 
landscapes of protein folding~\cite{shoemaker_wolynes_wang_pnas,plotkin_wang_wolynes_jcp}
 and glassy dynamics~\cite{wolynes_jphysfrance_1997}. 
Here, we study the effect of correlation in the discrete lattice model
(in the same spirit of the inherent correlation 
present in the Gaussian random field -- \equn~\ref{eq:field_correlation}), 
such that
\begin{equation} 
\label{eq:discrete_correlation}
    \avg{U_{j}U_{i}} = \varepsilon^2\exp \prth{-\frac{\sigma^2(j-i)^2}{2}}
\end{equation}
where $\sigma$ is now a measure of the spatial correlation on the lattice.
We derived the diffusion coefficient on this correlated potential using MFPT formalism,
\begin{equation} 
\label{eq:D_lattice_corr}
    \Deff = \bareD \exp \prth{-\beta ^{2}\varepsilon^{2}} 
            \sqbr{{1+{\rm erf} \prth{\displaystyle \frac{\beta\varepsilon}{2}\sqrt{1-\exp\prth{-\frac{\sigma^2}{2}}}}}}^{-1}
\end{equation}
This is a general expression of diffusion coefficient on a random lattice.
On an uncorrelated surface, \ie in the limit of  
$\sigma\to \infty$ the above expression reduces to \equn~\ref{eq:D_lattice_final}.

The diffusion coefficient on the continuous potential surface 
can be derived using adjoint operator technique~\cite{weiss_acp}.
Considering a reflecting boundary condition at $x=0$ and 
an absorbing boundary condition at $x=L$,
the $\tMFPT$ on the continuous potential surface $U(x)$ is obtained as,
\begin{equation} 
\label{eq:tMFPT_continuous} 
    \tMFPT = \frac{1}{\bareD} \int\limits_{0}^{\ L} e^{ \beta U(x)} dx 
                              \int\limits_{0}^{\ x} e^{-\beta U(y)} dy
\end{equation}
Using the same technique as used in the discrete model, 
the MFPT is then compared with that of a flat energy
surface with same boundary conditions. 
Assuming that the rough energy surface 
can be replaced by an effective flat energy surface, 
the effective diffusion coefficient is subsequently obtained from,
\begin{equation} 
\label{eq:D_mfpt_continuous_we}
    \Deff = \lim \limits_{L \to \infty } \frac{L^2}{2\tMFPT}  
\end{equation}
Under translation invariance, \equn~\ref{eq:tMFPT_continuous}
can be rewritten as,
\begin{equation}
\label{eq:tMFPT_continuous_deriv_01}
    \tMFPT = \frac{1}{\bareD} \int\limits_{0}^{\ L} d\xi \prth{L-\xi} 
             \avg{e^{\beta U(0) - \beta U(\xi)}}
\end{equation}
By substituting continuous limit of \equn~\ref{eq:discrete_correlation} into the above expression, 
we can calculate the MFPT using
\begin{equation}
\label{eq:tMFPT_continuous_deriv_02}
    \avg{e^{\beta U(0) - \beta U(\xi)}} = \exp \prth{\beta^2\avg{U(0)^2} - \beta^2\avg{U(0)U(\xi)}}
\end{equation}
and obtain the effective diffusion coefficient from \equn~\ref{eq:D_mfpt_continuous_we},
\begin{equation} 
\label{eq:D_field_final} 
    \Deff = \lim \limits_{L\to \infty }
            \frac{\displaystyle \bareD L^{2} \exp \prth{ -\beta^{2} \varepsilon^{2}}}
                 {\displaystyle 2 \int \limits_{0}^{\ L} d\xi \prth{L-\xi} 
                  \exp \sqbr{ -\beta^{2} \varepsilon^{2} \exp \prth{ -\sigma^2 \frac{\xi^{2}}{2} }}}
\end{equation}

We find that the diffusion coefficient 
for the correlated discrete potential 
-- \equn~\ref{eq:D_lattice_corr}
and the continuous potential (which is inherently correlated) 
-- \equn~\ref{eq:D_field_final} have similar implications.
In the limit of $\sigma\to 0$, 
\ie when the lattice is infinitely correlated,
$\Deff$ obtained from \equn~\ref{eq:D_lattice_corr} 
reduces to Zwanzig's form, $\Deffs{Z}$.
Similarly, the extension term in \equn~\ref{eq:D_field_final} 
also approaches $1$.
Therefore, Zwanzig's expression can be regarded 
as a limiting form in the case of infinitely long-range correlation. 
This explains the reason for the agreement observed in 
\fign~\ref{fig:diff_gaussian_field_result}.

\section{Apparent breakdown of ergodicity with increasing ruggedness}
\begin{figure}
\begin{center}
  \includegraphics[width=\figwidth\textwidth]{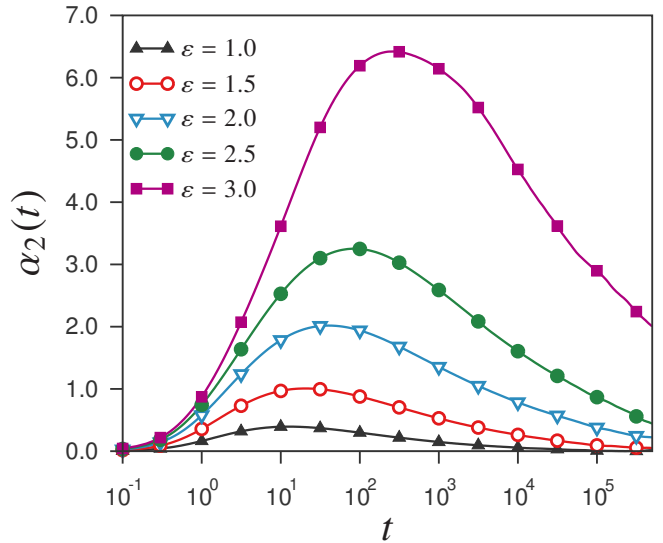}
\end{center}
\caption[]
      { \label{fig:diff_alpha2t}
        Non-Gaussian parameter $\alpha_{2}\prth{t}$ 
        for different randomness $(\varepsilon)$ in
        the continuous Gaussian random field. 
        The deviation from Gaussian behavior increases 
        with increasing $\varepsilon$.
        The deviation apparent breakdown of ergodicity in the system. 
        Hence reaching the diffusive limit becomes increasingly difficult.
      }
\end{figure}

\begin{figure}
\begin{center}
  \includegraphics[width=\figwidth\textwidth]{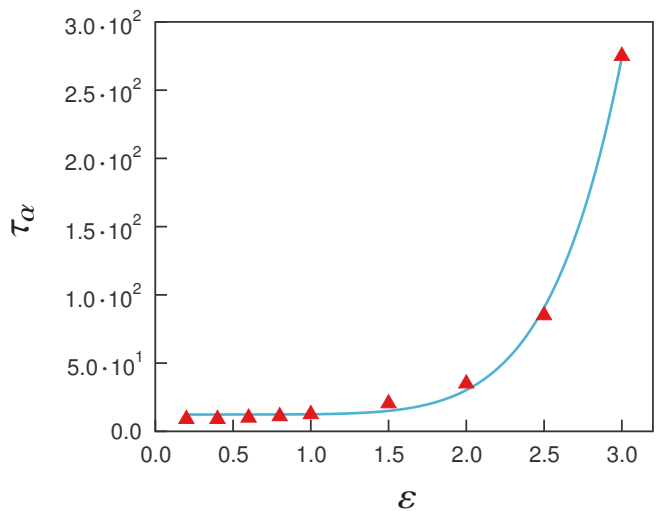}
\end{center}
\caption
      { \label{fig:diff_tau_scaling}
        Dependence of peak-maxima $\tau_{\alpha }$ of the non-Gaussian parameter, 
        on the magnitude of ruggedness parameter $\varepsilon$
        of the corresponding potential energy surface. 
        The red triangles are the results from simulations. 
        The solid blue line is the fitting to a power law,
        $\tau_{\alpha } =0.182\varepsilon^{6.618} + 12.363$ (see \equn~\ref{eq:tau_alpha_power_law}).
      }
\end{figure}

The present model provides a remarkably direct approach 
to study the relationship between diffusion and ergodicity.
At large ruggedness (large $\varepsilon $) 
our simulations tend to remain in the sub-diffusive regime.
Even for $\varepsilon > 3.0$ we could not reach the ergodic limit. 
The difficulty of reaching the ergodic limit
with increasing $\varepsilon $can be investigated 
using the non-Gaussian parameter, $\alpha _{2} \left(t\right)$.
It quantifies the deviation of the distribution of displacements 
from a Gaussian shape and is defined as~\cite{alpha_2_t_rahman},
\begin{equation} 
\label{eq:alpha2t} 
    \alpha_{2} (t) = \frac{\displaystyle \avg{\Delta x^{4} (t)}}
                          {\displaystyle \prth{1+\frac{2}{d}} \avg{\Delta x^{2} (t)}^2} - 1 
\end{equation}
where $d$ is the dimensionality of the system (in our case, $d=1$). 
For an ergodic system, the mean square displacement
of a particle increases linearly in time,
and the van Hove self-correlation function has a Gaussian shape.
In this case the non-Gaussian parameter is zero.
However, a non-zero value of non-Gaussian parameter
signifies a non-ergodic behavior.
The evolution of $\alpha _{2}\prth{t}$
with increasing $\varepsilon $ for the continuous potential model 
is shown in \fign~\ref{fig:diff_alpha2t}. 
With increasing $\varepsilon$,
the deviation from Gaussian behavior becomes more prominent. 
It indicates a very slow approach to diffusive behavior
for higher $\varepsilon $ that is expected to be re-established 
at very long times (which should scale with $\varepsilon$).
The peak maxima $\tau_{\alpha }$ gradually increases and shifts to longer time. 
The time $\tau_{\alpha }$ depends strongly on $\varepsilon$ and 
as shown in \fign~\ref{fig:diff_tau_scaling} can be fitted to a power law,
\begin{equation} 
\label{eq:tau_alpha_power_law}
    \tau_{\alpha } =a\varepsilon^{b} + c
\end{equation}
The obtained fitting parameters are 
$a=0.182, b=6.618 \text{ and } c=12.363$. 
This clearly indicates that beyond certain $\varepsilon $,
one needs exceedingly long time to reach the diffusive limit.

\section{Conclusion}
The present study demonstrates that even such 
apparently simple models of diffusion on a 
random Gaussian energy surface can reproduce many 
of the features observed in real experimental systems, 
such as crossover from ergodic to non-ergodic behavior, 
sharp rise in the peak of the non-Gaussian parameter 
and sub-diffusive dynamics. 
On the theoretical side, 
there are fundamental issues that need to be overcome. 
The breakdown of Zwanzig's elegant expression
was perhaps anticipated but was not clearly demonstrated earlier. 
We introduced an extension term that rectifies
Zwanzig's expression and we recommend that \equn~\ref{eq:D_lattice_final} 
be used instead of Zwanzig's expression 
for a random uncorrelated Gaussian surface. 
Similarly, \equn~\ref{eq:D_field_final} is the correct form 
to use for a Gaussian field. 
We discuss the role of spatial correlation 
in a random landscape, and show that
Zwanzig's expression is valid in the
asymptotic limit of infinitely correlated 
Gaussian random energy surface. 
The present models can be extended
to treat many interesting issues~\cite{stein_prob_stat_old} more quantitatively.
In a future work, we shall address a dynamic derivation 
of Rosenfeld scaling relation.

Our discussion is restricted to one-dimensional diffusion
and there seems to be no generalization to higher dimensions.
While one may conjecture that this provides an insight to the multi-dimensional surface,
the problem remains open for future investigation.
Several interesting phenomena might appear in higher dimensions.
Particularly, the walker should be able to avoid
the deep minima and maxima formed by the three sites as discussed above, 
due to presence of alternate paths that would avoid those barriers.
As a result, the mean field treatment of Zwanzig is expected to hold as dimension goes to infinity.
Recent studies~\cite{wang_spiliopoulos_2011, wang_spiliopoulos_2012} have focused 
on extending techniques of efficient importance sampling schemes for 
simulating rare events associated with higher dimensions.
Such multiple scale techniques would allow further analysis on this interesting problem.

\section*{Acknowledgments}
We dedicate this work to Professor Robert W. Zwanzig, 
a pioneer and giant in the area of statistical mechanics,  
who served as a mentor, directly to one of us (BB) 
and to many others through his highly insightful papers and clear writings.

We thank Dr. R. S. Singh for many helpful discussions and critical comments.
This work was supported in parts by grants from 
Board of Research in Nuclear Sciences (BRNS) and 
Department of Science and Technology (DST), India.
BB acknowledges support from  J. C. Bose Fellowship (DST).

\bibliography{manuscript}

\end{document}